\documentclass[prd,amsmath,amssymb,showpacs,showkeys,
               nofootinbib,eqsecnum,
               preprint,12pt,tightenlines,floatfix,
               a4paper,english, 
               ]{revtex4-1}

\usepackage{amsmath}
\usepackage{amssymb}
\usepackage{graphicx}
\usepackage{hhline}
\usepackage{mwe}
\usepackage{textcomp}
\usepackage{slashed}
\makeatletter
\usepackage{babel}

\newcommand{\beq}{\begin{equation}}
\newcommand{\eeq}{\end{equation}}
\newcommand{\beqa}{\begin{eqnarray}}
\newcommand{\eeqa}{\end{eqnarray}}
\newcommand{\bsubeqs}{\begin{subequations}}
\newcommand{\esubeqs}{\end{subequations}}
\newcommand{\nn}{\nonumber}

\newcommand{\pa}{\partial}

\newcommand{\half}{\textstyle{\frac{1}{2}}}

\def\id{\makebox[0.6ex][l]{$1$}{\rm l}}  

\begin{document}

\noindent Phys. Rev. D 96, 076007 (2017)
\hfill   arXiv:1703.10585
%
\newline\vspace*{3mm}

\title{Mass generation by a Lorentz-invariant gas of spacetime defects
       \vspace*{7mm}}

\author{F.R. Klinkhamer}
\email{frans.klinkhamer@kit.edu}
\affiliation{Institute for
Theoretical Physics, Karlsruhe Institute of
Technology (KIT), 76128 Karlsruhe, Germany\\}
\author{J.M. Queiruga}
\email{jose.queiruga@kit.edu}
\affiliation{Institute for
Theoretical Physics, Karlsruhe Institute of
Technology (KIT), 76128 Karlsruhe, Germany\\}
\affiliation{Institute for Nuclear Physics,
Karlsruhe Institute of Technology  (KIT),
Hermann-von-Helmholtz-Platz 1,
76344 Eggenstein-Leopoldshafen, Germany\\ \\}

\begin{abstract}
\vspace*{3mm}
We present a simple model of defects embedded in flat spacetime,
where the model is designed
to maintain Lorentz invariance over large length scales.
Even without remnant Lorentz violation,
there are still effects from these spacetime defects
on the propagation of physical fields, notably mass generation for scalars
and Dirac fermions.
\end{abstract}

\maketitle

\section{Introduction}
\label{sec:intro}

It has been argued that spacetime over small length
scales might have a nontrivial
structure~\cite{Wheeler1,Wheeler2,Hawking1,Hawking2,Hawking3}.
What the precise
nature of this small-scale ``structure'' would be is, however,
unclear.

It is known, for example, that static Swiss-cheese-type models
affect particle propagation and experimental data strongly
constrain the ``holes'' of such a classical
spacetime~\cite{Klink1,KS2008}.
As the static holes of such a model violate Lorentz invariance,
the above-mentioned bounds strongly constrain this type of
Lorentz violation;
see also the related discussion in Ref.~\cite{Collins-etal2004}.
The conclusion appears to be that, if spacetime somehow has a
small-scale structure, the underlying (quantum) theory manages
to keep Lorentz invariance to high precision.

For this reason, it may be of interest to investigate
toy models of spacetime-defects, where the models are designed
to maintain Lorentz invariance on average.
One class of such models involves pointlike defects,
as studied in
Refs.~\cite{KlinkhamerRupp2004,KlinkhamerRupp2005,Schreck-etal2013}
(see also Refs.~\cite{Hossen1,Hossen2}
for a general discussion).  In the present article,
we present one further toy model with pointlike defects
and study the induced modifications of the standard
particle propagation (``standard'' referring to the perfect
Minkowski spacetime without defects).

\section{Poisson distribution and Lorentz invariance}
\label{sec:Poisson}

It is a nontrivial issue to find
distributions of defects over four-dimensional Minkowski spacetime,
which preserve the Lorentz symmetry in the large. Let us assume,
for example, that the defects are distributed over a regular
hypercubic lattice
in one particular reference frame. Averaged over large scales, the distribution is homogeneous. But if we go to a Lorentz-boosted frame, the density of defects will increase in the direction of the boost, while remaining constant in the perpendicular directions. Apparently, the Lorentz symmetry is broken by having a preferred reference frame
in the original setup with a regular lattice.

Still, if the defects are distributed according to a Poisson process
(a ``sprinkling'' procedure), boosts do not break Lorentz invariance \cite{Dowker1,Bombelli,Henson}. The probability of finding $n$ defects in a four-dimensional volume $V_{4}$ is then given by
\beq
P_{n}\left(V_{4}\right)=
\frac{1}{n !}\,
\left(\rho_{d}\, V_{4}\right)^n \exp \left(-\rho_{d}\, V_{4} \right)
\label{Pois}\,.
\eeq
The parameter $\rho_{d}$ characterizes the distribution and corresponds to the average spacetime density of defects. Note that the Poisson process, for
constant parameter $\rho_{d}$, depends only on the four-dimensional volume of the region considered. This implies that the probability of finding $n$ defects contained in a region of volume $V_{4}$ is invariant under volume-preserving transformations. Since Lorentz transformations preserve
the spacetime volume, the sprinkling is Lorentz invariant. Phrased in a different way, the defect distribution from the Poisson process
has no built-in ``structure.'' If present, such a built-in structure
would be deformed by Lorentz contraction, just as for the
regular-lattice setup discussed above.

We see immediately from the Poisson distribution \eqref{Pois}
that, on average, the typical number of defects inside a region of volume $V_{4}$ is given by $\langle n\rangle_{V_{4}}=\rho_{d}\, V_{4}$ and
that the fluctuations of this number are of order $\sqrt{\rho_{d} V_{4}}$,
so that the relative fluctuations become irrelevant for large $V_{4}$.

\section{Effective model for a gas of spacetime defects}
\label{sec:Effective-model-gas-defects}

\subsection{General remarks}
\label{subsec:General-remarks}

The explicit calculation of physical observables in a theory
with finite-size spacetime defects
(corresponding to, e.g., soliton-type solutions~\cite{Klinkhamer2014})
is prohibitively difficult.
We can use, instead, a simple model with a gas of pointlike
defects~\cite{KlinkhamerRupp2004,Schreck-etal2013}.

In the new model presented here, the spacetime defects are represented
by randomly-positioned
delta functions in a classical background Minkowski spacetime,
where the delta functions are coupled to a ``mediator''
real scalar field $\sigma(x)$ with random charges $\epsilon_{n}\in\lbrace -1,1\rbrace$ and a coupling constant $\lambda$.
The charges of the individual defects are randomly chosen
with probability 1/2 to get charge $+1$
and  probability 1/2 to get charge $-1$,
so that the  average
charge vanishes over a large enough spacetime volume.
The mediator field $\sigma(x)$ is also coupled to three ``physical'' fields:
a massless real scalar field $\phi(x)$ with a nonderivative quartic-coupling
term, a massless Dirac fermion field $\psi_{1}(x)$
with a Yukawa coupling term,
and a massless Dirac fermion field $\psi_{2}(x)$ with a
nonrenormalizable Yukawa-type coupling term.

In short, a nontrivial spacetime with defects is modeled by a perfect
classical Minkowski spacetime and an action with delta functions
coupled to a real scalar field $\sigma(x)$. In turn, this mediator
field $\sigma(x)$ is coupled to physical fields
$\phi(x)$, $\psi_{1}(x)$, and $\psi_{2}(x)$,
where all fields propagate over Minkowski spacetime.

\subsection{Massless mediator field}
\label{subsec:Massless-mediator}

The following effective action is considered:
\bsubeqs\label{eq:action-gas-delta-functions-epsilon-n-Ppls-Pminus}
\beqa
\label{eq:action-gas-delta-functions}
S_{\text{eff}}&=&
- \int_{\mathbb{R}^4}d^4 x\,\bigg(
\half\,\eta^{\mu\nu}\,\pa_\mu\phi\,\pa_\nu\phi
+i \bar{\psi_{1}} \gamma^\mu\pa_\mu\psi_{1}
+i \bar{\psi_{2}} \gamma^\mu\pa_\mu\psi_{2}
\nn\\[1mm]&&
+\half\,\eta^{\mu\nu}\, \pa_\mu\sigma\,\pa_\nu\sigma
+\lambda\, \sigma\,
\Big[\sum_{n=1}^\infty \epsilon_{n}\, \delta^{(4)}(x-x_{n})\Big]_{\{x_{n}\}\;\text{from\;Poisson}}
\nn\\[1mm]&&
+g_{s}\, \sigma^2\,\phi^2
+g_{f,1}\,\sigma\,\bar{\psi_{1}}\psi_{1}
+g_{f,2}\,\lambda\,\sigma^2\,\bar{\psi_{2}}\psi_{2}
\bigg)\,,
\eeqa
\beqa
\label{eq:epsilon-n-Pplus-Pminus}
\epsilon_{n} &\in&\lbrace -1,\,+1\rbrace\,,
\;\;\text{with}\;\;
P_{-1}=P_{+1}=1/2 \,,
\eeqa
\esubeqs
where the Minkowski metric is given by
$\eta_{\mu\nu}\equiv [\text{diag}(-1,\,1,\,1,\,1)]_{\mu\nu}$
for standard Cartesian coordinates $x^\mu$.
The dimensionless quartic scalar coupling constant $g_s$
is taken to be positive,
so that the scalar potential is bounded from below.
Throughout, we use natural units with $\hbar=1=c$.

In the action (\ref{eq:action-gas-delta-functions}), the cores of the
spacetime defects are modeled by Dirac delta functions centered at the points $x_{1},x_{2},...$ of Minkowski spacetime.
As discussed in Sec.~\ref{sec:Poisson}, these points are distributed according to a Poisson process (sprinkling), in order to preserve the Lorentz symmetry.
The long-range effects of the defect cores are modeled by
a real scalar field $\sigma$ with coupling strength $\lambda$ and
random charges $\epsilon_{n}\in\lbrace -1,1\rbrace$.
With equal probabilities \eqref{eq:epsilon-n-Pplus-Pminus}
for having a positive and a negative charge
(no correlation between the different defects),
the average charge vanishes asymptotically,
\beq
\label{eq:average-charge-vanishing}
\lim_{N\to\infty}\,\frac{1}{N}\,\sum_{n=1}^{N}\, \epsilon_{n} = 0\,.
\eeq
The scalar field $\sigma(x)$ mediates between the defect cores
at random positions $x=x_{n}$
and the physical fields $\phi(x)$ and $\psi_{a}(x)$, for $a=1,\,2$.

The massless scalar fields $\phi$ and $\sigma$ in \eqref{eq:action-gas-delta-functions}
have mass dimension $1$, the massless fermionic fields $\psi_{a}$ have mass dimension $3/2$, the coupling constant $\lambda$ has mass dimension $-1$,
and the couplings $g_{s}$ and $g_{f,a}$ are dimensionless.
As mentioned in Sec.~\ref{subsec:General-remarks},
the idea behind the action
\eqref{eq:action-gas-delta-functions-epsilon-n-Ppls-Pminus}
is that a nontrivial spacetime (manifold or not) is modeled by
delta functions located at the
spacetime points $x_{n}$ of the Minkowski manifold
and by a mediator field $\sigma(x)$. The mediator field $\sigma(x)$
is coupled to the delta functions and to additional physical fields $\phi(x)$ and $\psi_{a}(x)$, with all fields propagating over classical
Minkowski spacetime and the interaction terms given by
the last three terms of the integrand of (\ref{eq:action-gas-delta-functions}).

In order to recover the standard perturbative results
by use of Feynman diagrams, the interactions terms of \eqref{eq:action-gas-delta-functions}
essentially need to be ``turned off'' in the asymptotic
regions~\cite{Feynman1949}.
This can be done by making the couplings
in \eqref{eq:action-gas-delta-functions} spacetime dependent,
\beqa
&&
\{\lambda(x),\, g_{s}(x),\, g_{f,a}(x) \}
=
\left\{
\begin{array}{ll}
\{\overline{\lambda},\, \overline{g}_{s},\, \overline{g}_{f,a} \}\,,\quad
& \text{for}\;\;x\in V_{4,\,\text{cutoff}}\,,    \\[2mm]
\{0,\, 0,\, 0 \}\,,
&\text{otherwise}\,,
\end{array}
\right.
\label{eq:coupling-constants-turn-off}
\eeqa
where the barred quantities on the right-hand side are truly constant.
The spacetime volume
$V_{4,\,\text{cutoff}}$ in \eqref{eq:coupling-constants-turn-off}
is taken to be suitably large
and the behavior in the transition region can be adequately
smoothed.
In the following, we will keep this spacetime dependence of the couplings
$\{\lambda,\, g_{s},\, g_{f,a} \}$ implicit.
Incidentally, restricting the sum in \eqref{eq:action-gas-delta-functions}
to the finite volume $V_{4,\,\text{cutoff}}$ makes this sum well behaved,
with a finite number $N$ of defects.

The classical solution for the mediator field $\sigma$ can be easily obtained for $g_{s}=g_{f,a}=0$,
\beq\label{eq:sigma-equation}
\frac{\delta S_{\text{eff}}}{\delta\sigma}\,
\Bigg|^{(g_{s}=g_{f,a}=0)}
=0	
\;\;\Rightarrow\;\;
\square\,\sigma(x) =
\lambda\sum_{n} \epsilon_{n}\, \delta^{(4)}(x-x_{n})\,,
\eeq
with the flat-spacetime
d'Alembert operator $\square\equiv \partial^\mu \partial_\mu$.
The solution is given by
\beq
\sigma(x)=\sigma_0(x)
+\lambda\sum_n\epsilon_{n}\int d^4x'\,  G_{0}(x,x')\, \delta^{(4)}(x-x_{n})\,,
\eeq
where $\sigma_0(x)$ is the free solution [corresponding to the homogeneous equation (\ref{eq:sigma-equation}) with $\lambda=0$]
and $G_{0}(x,x')$ is a Green's function of
the d'Alembert operator $\square$,
\beq
G_{0}(x,x')=\frac{1}{4\pi^2\, \vert x-x'\vert^2}\,.
\eeq
Explicitly, the solution of \eqref{eq:sigma-equation} takes the following form:
\beq
\sigma(x)=\sigma_0(x)
+\lambda\,\sum_{n}\frac{\epsilon_{n}}{4\pi^2\, \vert x-x_{n}\vert^2}
\equiv \sigma_0(x) +\sigma_{1}(x)\,.
\label{totals}
\eeq

Let us now determine the two-point function for $\sigma$. In the quantization procedure, the free part of $\sigma$ can be split in positive and negative frequency modes as usual,
\beq
\sigma_0(x)=\int \frac{d^3 p}{(2\pi)^3}\frac{1}{\sqrt{2\,\omega_{p}}}
\left(a_p\,  e^{-ipx}+a_p^\dagger\,  e^{ipx}\right)\,,
\label{mode}
\eeq
with $p_0 = \omega_{p} \equiv \sqrt{p_1^2+p_2^2+p_3^2}$.
If we define
\beq
j(x)\equiv \lambda\sum_{n}\epsilon_{n}\, \delta^{(4)}(x-x_{n})\,,
\eeq
the Fourier transform takes a simple form
\beq
j(p)=\int d^4 x\, e^{ipx}\, j(x)=\lambda\sum_{n}\epsilon_{n}\,  e^{ipx_{n}}\,.
\label{fou}
\eeq

With the help of (\ref{fou}), we can expand the correction term in (\ref{totals}) as follows:
\beq
\sigma_{1}(x)=\lambda\,
\int \frac{d^3 p}{(2\pi)^3}\left(\frac{  \sum_{n}\epsilon_{n} \, e^{ipx_{n}}\, e^{-ipx}}{2\,\omega_{p}}\,\id +\text{H.c.}\right)\label{corsig}\,,
\eeq
where $\id $ is the identity operator. The only nonvanishing contributions to the  two-point function come from terms proportional to $\langle 0 \vert\, a_p\, a_q^\dagger \,\vert  0\rangle=(2\pi^3)\, \delta^{(3)}(p-q)$ or proportional to
$\langle 0 \vert\, \id ^2 \,\vert  0\rangle=1$.
The final result is given by
\beqa
 &&
 \langle 0 \vert \sigma(x)\sigma(y)\vert  0\rangle=
 \int \frac{d^3p}{(2\pi)^3}\frac{e^{-ip(x-y)}}{2\,\omega_{p}}
 +4\,\lambda^2\, \sum_{m,n}\epsilon_{n}\,\epsilon_{m} \int \frac{d^3p}{(2\pi)^3}\frac{e^{-ip(x-x_{n})}}{2\,\omega_{p}} \int \frac{d^3q}{(2\pi)^3}\frac{e^{-iq(x_{m}-y)}}{2\omega_q}\label{2p}\,.
 \nn\\[1mm]&&
\eeqa
Taking into account that the first term on
the right-hand side of (\ref{2p}) corresponds to the free two-point function [denoted by $\Delta_{0}(x-y)$ as usual] we get
\beqa
&&
 \langle 0 \vert \sigma(x)\sigma(y)\vert  0\rangle= \Delta_{0}(x-y)
 +4\,\lambda^2\, \sum_{m,n}\epsilon_{m}\,\epsilon_{n} \, \Delta_{0}(x-x_{n})\, \Delta_{0}(x_{m}-y)\label{2pointnon}\,.
\eeqa

The expression \eqref{2pointnon} has a simple interpretation:
the amplitude for the scalar field $\sigma$ to propagate
from a spacetime point $x$ to a spacetime point $y$ is given by the free amplitude plus all possible products of
the free amplitude
of particle propagation from $x$ to the position of a defect $x_{n}$
times the free amplitude of particle propagation
from another defect $x_{m}$ to $y$. In the random-phase approximation
(see Appendix \ref{app:Random-phase-approximation}),
all cross terms joining different defects in \eqref{2pointnon}
are subdominant.
The expression (\ref{2pointnon}) then simplifies to
\beqa
&&
\langle 0 \vert \sigma(x)\sigma(y)\vert  0\rangle
\approx
\Delta_{0}(x-y)
+4\,\lambda^2\, \sum_{n} \Delta_{0}(x-x_{n})\, \Delta_{0}(x_{n}-y)\label{2pn}\,.
\eeqa
This full tree-level propagator contains the free propagator and the sum of all possible insertions of a single defect.

\subsection{Massive mediator field}
\label{subsec:Massive-mediator}

For completeness,  we also consider the case
where the mediator field has a nonzero initial mass $m_0$.
The massive version of
the original action (\ref{eq:action-gas-delta-functions})
takes the following form:
\beqa
S_{\text{eff},\,m_0}&=&
- \int_{\mathbb{R}^4}d^4 x \,
\Bigg(
\half\,\eta^{\mu\nu}\,\pa_\mu\phi\,\pa_\nu\phi
+i \bar{\psi_{1}} \gamma^\mu\pa_\mu\psi_{1}
+i \bar{\psi_{2}} \gamma^\mu\pa_\mu\psi_{2}
+\half\,\eta^{\mu\nu}\, \pa_\mu\sigma\,\pa_\nu\sigma
+\half\,m_0^2\,\sigma^2
\nn\\[1mm]&&
+\lambda\,\sigma\sum_{n=1}^\infty\epsilon_{n}\, \delta^{(4)}(x-x_{n})
+g_{s}\,\sigma^2\,\phi^2
+g_{f,1} \,\sigma\,\bar{\psi_{1}}\,\psi_{1}
+g_{f,2}\,\lambda\,\sigma^2\,\bar{\psi_{2}}\psi_{2}
\Bigg)\,,
\label{gasdeltamassive}
\eeqa
with the physical fields $\phi$ and $\psi_{a}$ still being massless,
as long as interactions are neglected.

As  before, the classical equation for $\sigma$ can be written as
\beq
\left(\square-m_0^2\right) \sigma(x) =
\lambda\sum_n\epsilon_{n}\,\delta^{(4)}(x-x_{n})\,,
\eeq
again setting $g_{s}=g_{f,a}=0$.
The complete solution of this equation can be split
in two parts,
\beq
\sigma(x)=\sigma_0(x)
+\lambda\sum_n\epsilon_{n}\int d^4x'\, \widetilde{G}_{0}(x,x')\,\delta^{(4)}(x-x_{n})\,,
\eeq
where, now, $\sigma_0(x)$ is the solution of the homogeneous equation $\left(\square-m_0^2\right)\sigma_0(x)=0$ and $\widetilde{G}_{0}(x,x')$ is a Green's function of the operator $\square-m_0^2$,
\beq
\widetilde{G}_{0}(x,x')=
\frac{m_0 \, \mathcal{K}_{1}(m_0\vert x-x' \vert)}{4\pi^2 \vert x-x' \vert}\,,
\eeq
with $\mathcal{K}_{\nu}(z)$ the modified Bessel function
of the second kind~\cite{AbramowitzStegun1965}.

A calculation similar to the one of Sec.~\ref{subsec:Massless-mediator}      gives the following result for the two-point function:
\beqa
&&
 \langle 0 \vert \sigma(x)\sigma(y)\vert  0\rangle_{m_0}
 = \widetilde{\Delta}_{0} (x-y)
 +4\,\lambda^2\, \sum_{n,m} \epsilon_{n}\,\epsilon_{m}\,\widetilde{\Delta}_{0}(x-x_{n})\,
 \widetilde{\Delta}_{0}(x_{m}-y)\,,
 \label{2pm}
\eeqa
where $\widetilde{\Delta}_{0}(x-y)$ is the free massive propagator.
In the random-phase approximation (see Appendix \ref{app:Random-phase-approximation}),
the expression (\ref{2pm}) simplifies to
\beqa
&&
\langle 0 \vert \sigma(x)\sigma(y)\vert  0\rangle_{m_0}
\approx
\widetilde{\Delta}_{0} (x-y)
+4\,\lambda^2\, \sum_{n}\widetilde{\Delta}_{0}(x-x_{n})\, \widetilde{\Delta}_{0}(x_{n}-y)
\label{2pmr}\,,
\eeqa
which has the same structure as expression \eqref{2pn}
for the massless-mediator case.

\section{Mass generation for the mediator field}
\label{sec:Mass-generation-mediator-field}

As found in Sec.~\ref{sec:Effective-model-gas-defects},
the interaction of the mediator field $\sigma(x)$ with the delta functions
leads to a nontrivial modification of the $\sigma$ propagator.
Let us focus on the initially massless case given by the action (\ref{eq:action-gas-delta-functions}). The propagator for the mediator field (\ref{2pn}) can then be rewritten as follows:
\beq
\Delta(x,\,y)
\approx
\Delta_0(x-y)+4\,\lambda^2\,\Delta_{1}(x,\,y)\label{propmas}\,,
\eeq
where $\Delta_0(x-y)$ corresponds to the free propagator and $\Delta_{1}(x,\,y)$ to the correction from the interactions with
the delta functions (corresponding to the defect cores).
We will see that this last term $\Delta_{1}$ generates a nonzero mass for the $\sigma$ field.

If a nonzero mass $m_{\sigma}$ is indeed generated,
then the following equation must hold:
\beq
\left(\square_x-m_{\sigma}^2\right)\Delta(x,\,y)=-\delta^{(4)}(x-y)\,,
\label{green}
\eeq
for $m_{\sigma}^2\neq 0$. After inserting the propagator (\ref{propmas}) in (\ref{green}), we obtain to order $\lambda^2$
 \beqa
 &&
 4\,\lambda^2\, \sum_{n} \delta^{(4)}(x-x_{n})\Delta_0(x_{n}-y)
 -m_{\sigma}^2 \Delta_0(x-y)
 +\mathcal{O}\left[\lambda^4\,\rho_{d}^2\,\Delta_{1}(x,\,y)\right]
 =0\,.\label{green1}
 \eeqa
It is still not easy to interpret the first term on the left-hand side of (\ref{green1}). To do so, we can use the fact that the points $x_{n}$ are distributed according to a Poisson process as discussed in
Sec.~\ref{sec:Poisson}.
According to (\ref{Pois}),  the number of defects grows with the spacetime volume, $dN \propto d^4x$, where the proportionality factor is given by the density parameter $\rho_{d}$. Furthermore, the distribution is assumed to be dense and, therefore, the characteristic distance between defects, $l_{d}\equiv\rho_{d}^{-1/4}$, is assumed to be small
compared to the typical wavelengths of the fields considered.
This allows us to approximate the sum in  (\ref{green1}) by an integral,
\beq
\sum_{n}\rightarrow \rho_{d}\int_{\mathbb{R}^{4}}d^4x\,.\label{assump}
\eeq
Applying (\ref{assump}) to (\ref{green1}) we obtain the result
\beq
m_{\sigma}^2=4\,\lambda^2\,\, \rho_{d}\,,
\label{massb}
\eeq
in terms of the defect density $\rho_{d}$ from \eqref{Pois}
and the coupling constant $\lambda$ from
\eqref{eq:action-gas-delta-functions}.

The first corrections to the mass-square (\ref{massb}) will appear as loop corrections involving the dimensionless couplings $g_{s}$ and $g_{f,a}$.

We conclude that, as a result of the interactions with the delta functions, the mediator field $\sigma$ has acquired a mass. This result can be confirmed by working in the momentum-space representation.
Start from the full propagator (\ref{2pn})
and take (\ref{assump}) into account,
\beqa
&&
 \langle 0 \vert \sigma(x)\sigma(y)\vert  0\rangle
\approx
 \Delta_{0}(x-y)
 +4\,\lambda^2\, \rho_{d}\int d^4z\, \Delta_{0} (x-z)\,\Delta_{0} (z-y)\label{2pnn}\,.
\eeqa
After shifting the $z$ variable (in order to make explicit the dependence of the two-point function on the difference $x-y$),
we can rewrite (\ref{2pnn}) in momentum space as follows:
\beq
G(p)=\frac{1}{p^2+i\epsilon}
+4\,\lambda^2\, \rho_{d} \int d^4 z\,
\Delta_{0} (z)\,\frac{e^{ipz}}{p^2+i\epsilon}\,.
\eeq
Integration with respect to $z$ then gives
\beq\label{eq:G-sigma-field}
G(p)=\frac{1}{p^2+i\epsilon}
-\frac{1}{p^2+i\epsilon}\,
4\,\lambda^2\, \rho_{d} \,\frac{1}{p^2+i\epsilon}\,.
\eeq
The expression \eqref{eq:G-sigma-field}
can be rewritten to quadratic order in $\lambda$ as follows:
\beq
G(p)=\frac{1}{p^2-4\,\lambda^2\,\rho_{d}+i\epsilon}
+\mathcal{O}
\left[\lambda^4\, \rho_{d}^2\,(p^2+i\epsilon)^{-3}\right]\,.
\eeq
The mass-square term \eqref{massb}
for $\sigma$ appears in this representation as a pole in the momentum-space propagator. In the limit $\lambda\rightarrow 0$ (no long-range effects of the defect cores) and/or in the limit $\rho_{d}\rightarrow 0$ (vanishing density of defect cores), the mediator field remains massless.

For the initially massive case (\ref{gasdeltamassive}), the pole in
the $\sigma$-propagator is already shifted by the mass-square $m_0^2$
and an effective mass-square $m_{\sigma}^2=m_0^2+4\,\lambda^2\,\rho_{d}$
is obtained.

\section{Physical fields}
\label{sec:Physical-fields}

\subsection{Setup}
\label{subsec:Setup}

The massless physical fields $\phi$ and $\psi_{a}$ only ``feel'' the presence of the defect cores by interaction with the mediator field $\sigma$.
From now on, we work with the effective action obtained from
(\ref{eq:action-gas-delta-functions})
for a vanishing initial $\sigma$ mass.
We can write the obtained effective action
at order $\lambda^2$ as follows
 \beqa
S_{\text{eff},\,\lambda^2}&=&
-  \int_{\mathbb{R}^4}d^4 x\,\Big(
\half\,\pa_\mu\, \phi\pa^\mu\phi
+i \bar{\psi_{1}} \gamma^\mu\pa_\mu\psi_{1}
+i \bar{\psi_{2}} \gamma^\mu\pa_\mu\psi_{2}
+\half\,\pa_\mu\sigma\, \pa^\mu\sigma
+2\lambda^2\,  \rho_{d}\, \sigma^2
\nn\\[1mm]&&
+g_{s}\, \sigma^2\, \phi^2
+g_{f,1}\,  \sigma\, \bar{\psi_{1}}\psi_{1}
+g_{f,2}\,\lambda\,\sigma^2\,\bar{\psi_{2}}\psi_{2}
+\mathcal{O}\left[\lambda^4\,\rho_{d}^2\right] \Big) \,,
\label{gasdelta11}
\eeqa
where the delta functions have produced a mass term for the $\sigma$ field
as calculated in Sec.~\ref{sec:Mass-generation-mediator-field}.

\subsection{Physical scalar field $\phi$}
\label{subsec:Scalar-field-phi}

\begin{figure}[t]
\includegraphics[width=0.75\linewidth]{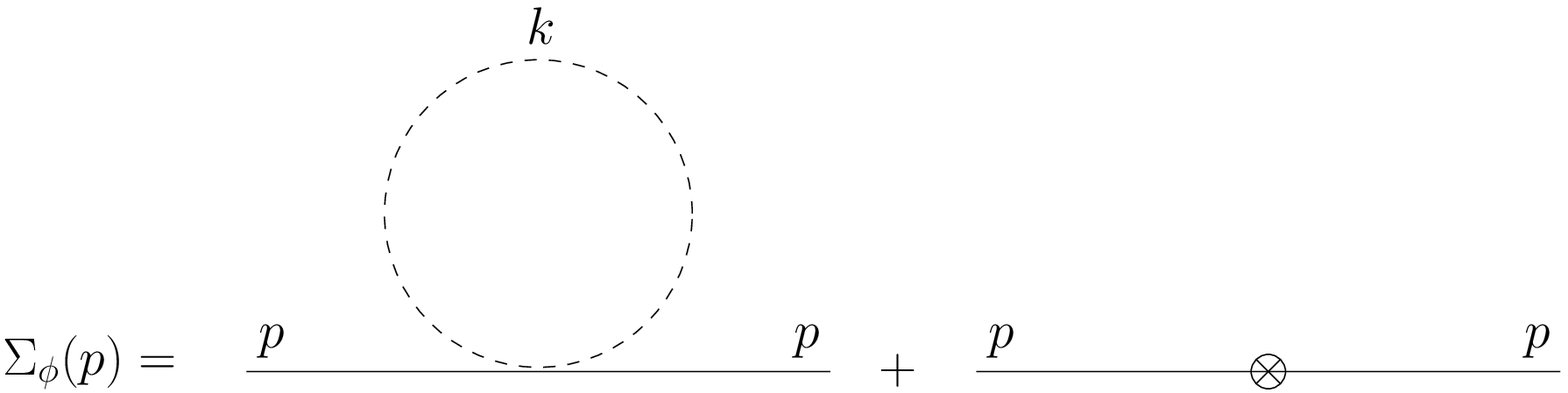}
\caption{\label{fig:self-energy-phi}
 Self-energy contribution for the physical scalar $\phi$.
 The dashed propagator corresponds to the mediator scalar field $\sigma$
 and the counterterm is given in Fig.~\ref{fig:counterterm-phi}.}
\vspace*{15mm}
\includegraphics[width=0.45\linewidth]{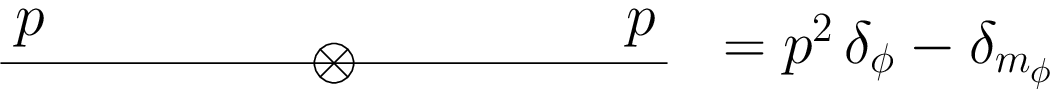}
\caption{\label{fig:counterterm-phi}
 Counterterm for Fig.~\ref{fig:self-energy-phi}.}
\end{figure}

We now ask what happens to the massless physical field $\phi$
by its interaction with the mediator field $\sigma$. The self-energy for the scalar field is given by Fig.~\ref{fig:self-energy-phi}.
\noindent
With appropriate regularization,
the last term in Fig.~\ref{fig:self-energy-phi} corresponds the counterterm of Fig.~\ref{fig:counterterm-phi}. This counterterm cancels the divergence coming from the one-loop integral.

Consider the Pauli--Villars (PV) regularization method~\cite{PauliVillars1949,ItzyksonZuber1980}.
For the 1-loop diagram of Fig.~\ref{fig:self-energy-phi}, the divergent integral is
 \beq\label{eq:divergent-integral-scalar}
 I(m^2_{\sigma}) \equiv
g_{s}\, \int \frac{d^4 k_E}{(2\pi)^4\,}\frac{1}{k^2_E+m_\sigma^2} \,,
\eeq
where $k_E$ is the Euclidean momentum.
We now define the PV-regularized integral by
 \beq\label{eq:divergent-integral-scalar-PV}
 I_\text{PV}(m^2_{\sigma})= I(m^2_{\sigma})-I(\Lambda^2)
 -(m_{\sigma}^2-\Lambda^2)\,I'(\Lambda^2)\,,
 \eeq
with the PV regulator mass $\Lambda$
and $I'(x) \equiv d I(x)/d x$.
Evaluating \eqref{eq:divergent-integral-scalar-PV} gives
 \beq\label{eq:divergent-integral-scalar-PV-result}
  I_\text{PV}(m^2_{\sigma})=
  -g_{s}\,\frac{m_{\sigma}^2}{(4\pi)^2}+g_{s}\,\frac{\Lambda^2}{(4\pi)^2}
  \left(1-\log\frac{\Lambda^2}{m_{\sigma}^2}\right)\,.
 \eeq
The counterterm of Fig.~\ref{fig:counterterm-phi} cancels exactly the
$\Lambda$ terms of (\ref{eq:divergent-integral-scalar-PV-result}),
\begin{subequations}\label{phi-counterterms-PV-reg}
\begin{eqnarray}
\delta_\phi &=& 0   \,,
\\[2mm]
\delta_{m_\phi}&=&
g_{s}\,\frac{\Lambda^2}{(4\pi)^2}
  \left(1-\log\frac{\Lambda^2}{m_{\sigma}^2}\right) \,.
\end{eqnarray}
\end{subequations}
As a result, a nonzero mass for $\phi$ is generated at one-loop level,
\beq
M_\phi^2\,\Big|^\text{(PV\,reg.)}
=\lim_{p\rightarrow 0} \,\Sigma_{\phi}(p)\,\Big|^\text{(PV\,reg.)}
=g_{s}\,\frac{m_{\sigma}^2}{(4\pi)^2}\,,
\label{eq:massphi-PV-reg}
\eeq
with $m_{\sigma}^2=4\,\lambda^2\,\,\rho_{d}$ from \eqref{massb}.
The point-splitting and
dimensional-regularization methods~\cite{ItzyksonZuber1980} give a
similar result for the generated scalar mass-square, with the same
parametric dependence $g_{s}\,m_{\sigma}^2$.

The generated mass-square for the scalar field
as given by (\ref{eq:massphi-PV-reg}) depends linearly on both $\rho_{d}$ and $\lambda^2$. This implies that, in order to give a nonzero mass to the scalar field,  both the presence of defect
cores ($\rho_{d}\neq 0$) and the interaction of defect cores with the mediator field ($\lambda\neq0$) are essential.

Two general remarks are in order. First,
the underlying nontrivial spacetime produces not only
the model \eqref{eq:action-gas-delta-functions-epsilon-n-Ppls-Pminus}
but also the required counterterms such as \eqref{phi-counterterms-PV-reg}.
If a single energy scale $E_\text{foam}$ (equal or not equal to the
Planck energy $E_{P}\equiv G^{-1/2}$) sets the parameters
$\lambda\sim 1/E_\text{foam}$ and $\rho_{d} \sim (E_\text{foam})^4$,
then it is also to be expected that $\Lambda \sim E_\text{foam}$,
and the Pauli--Villars ``regularization'' is no longer a mere mathematical device but is rooted in physical reality.
The generated mass-square \eqref{eq:massphi-PV-reg} can be very much
smaller than $(E_\text{foam})^2$ if $g_{s} \ll 1$.

Second, the following simple question arises:
is it not possible that the generated mass-square
\eqref{eq:massphi-PV-reg} gets absorbed
in the square of the renormalized mass
and that, thereby, the effects from our spacetime defects become
invisible? In general, this is certainly possible, but not for the
setup of the theory as outlined in Sec.~\ref{sec:Effective-model-gas-defects}.
Specifically, the coupling $\lambda$ is taken to vanish far out,
according to \eqref{eq:coupling-constants-turn-off}.
This means that the constant renormalized mass
relevant to the infinite-volume
perfect Minkowski spacetime is unaffected by spacetime defects, as their
effects ($\overline{\lambda}\ne 0$)
are confined to the finite volume $V_{4,\,\text{cutoff}}$.

\subsection{Physical fermion field $\psi_{1}$}
\label{subsec:Fermion-field-psi1}

\begin{figure}[t]
\includegraphics[width=0.75\linewidth]{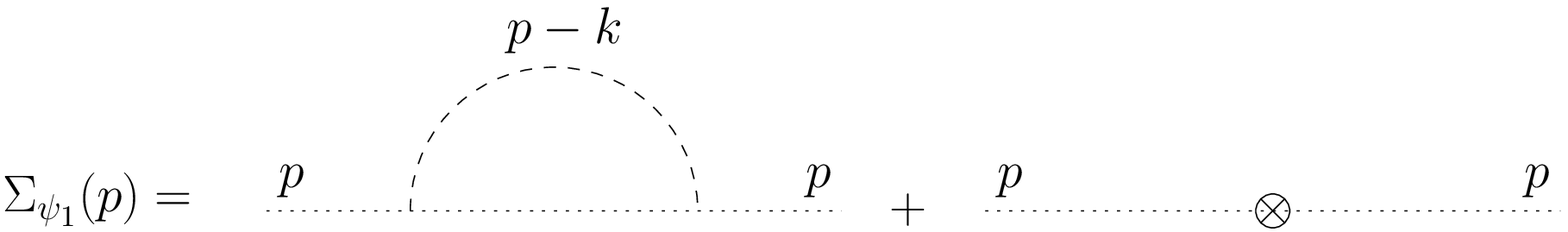}
\caption{\label{fig:self-energy-psi1}
Self-energy contribution for the physical fermion $\psi_{1}$.
The dashed propagator corresponds to the mediator scalar field $\sigma$
and the counterterm is given in Fig.~\ref{fig:counterterm-psi1}.}
\vspace*{15mm}
\includegraphics[width=0.45\linewidth]{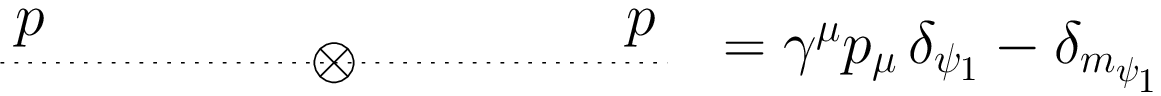}
\caption{\label{fig:counterterm-psi1}
 Counterterm for Fig.~\ref{fig:self-energy-psi1}.}
\end{figure}

The interaction of the massless fermionic field $\psi_{1}$ with $\sigma$ gives rise to different effects compared to those of the scalar field $\phi$.
The self-energy for the fermion including the counterterm is given by Figs.~\ref{fig:self-energy-psi1} and \ref{fig:counterterm-psi1}.
\vspace{0cm}
\noindent

Consider again Pauli--Villars regularization.
The unsubtracted regularized integral from the 1-loop diagram of Fig.~\ref{fig:self-energy-psi1}
gives
\beq
 \Sigma_{\psi_{1}}(p)\,\Big|^\text{(PV\,reg.,\;unsubtr.)}=
 -\frac{g_{f,1}^2}{16\pi^2}\,\slashed{p}\,
 \log\left(\frac{\Lambda^2}{m^2_{\sigma}}\right)
-\frac{g_{f,1}^2}{8\pi^2}\,\slashed{p}\,
\int_0^1dz\, z\,\log\left(\frac{m_{\sigma}^2}{m_{\sigma}^2-p^2(1-z)} \right) \,,
\eeq
with the Pauli--Villars regulator  $\Lambda$
and the Feynman slash notation $\slashed{p}\equiv \gamma^\mu p_\mu$.
The required counterterm is then
\begin{subequations}
\begin{eqnarray}
\delta_{\psi_{1}}&=&\frac{g_{f,1}^2}{16\pi^2}\,\slashed{p}\,
\log\left(\frac{\Lambda^2}{m^2_{\sigma}}\right)  \,,
\\[2mm]
\delta_{m_{\psi_{1}}}&=&0  \,.
\end{eqnarray}
\end{subequations}
This results in
\beq\label{eq:massfer-PV-reg}
M_{\psi_{1}}\,\Big|^\text{(PV\,reg.)}
=\lim_{p\rightarrow 0} \, \Sigma_{\psi_{1}}(p)\,\Big|^\text{(PV\,reg.)}
=0\,.
\eeq
The point-splitting and dimensional-regularization methods
also give a vanishing generated mass for $\psi_{1}$.

Hence, the fermion $\psi_{1}$ does not get a mass due to the interaction with the mediator field $\sigma$, at least at the 1-loop level and under the
assumption that $\sigma$ does not acquire a vacuum expectation value.
Expanding on this last point, the effective
action \eqref{gasdelta11} is invariant under the following
axial transformation:
\bsubeqs\label{eq:axial-transformations}
\beqa
\label{eq:axial-transformations-psi}
\psi_{1}(x)&\to& \exp[i(\pi/2)\gamma_5]\,\psi_{1}(x)\,,
\\[2mm]
\label{eq:axial-transformations-phi}
\sigma(x) &\to& -\sigma(x)\,.
\eeqa
\esubeqs
If unbroken, the axial symmetry \eqref{eq:axial-transformations}
of the effective action \eqref{gasdelta11}
rules out a direct (or generated) mass term $M \bar{\psi_{1}}\psi_{1}$.

\subsection{Physical fermion field $\psi_{2}$}
\label{subsec:Fermion-field-psi2}

The interaction of the massless fermionic field $\psi_{2}$ with $\sigma$
differs from that of $\psi_{1}$ with $\sigma$ and
is, in fact, similar to the interaction of
the scalar field $\phi$ with $\sigma$.

\begin{figure}[t]
\includegraphics[width=0.75\linewidth]{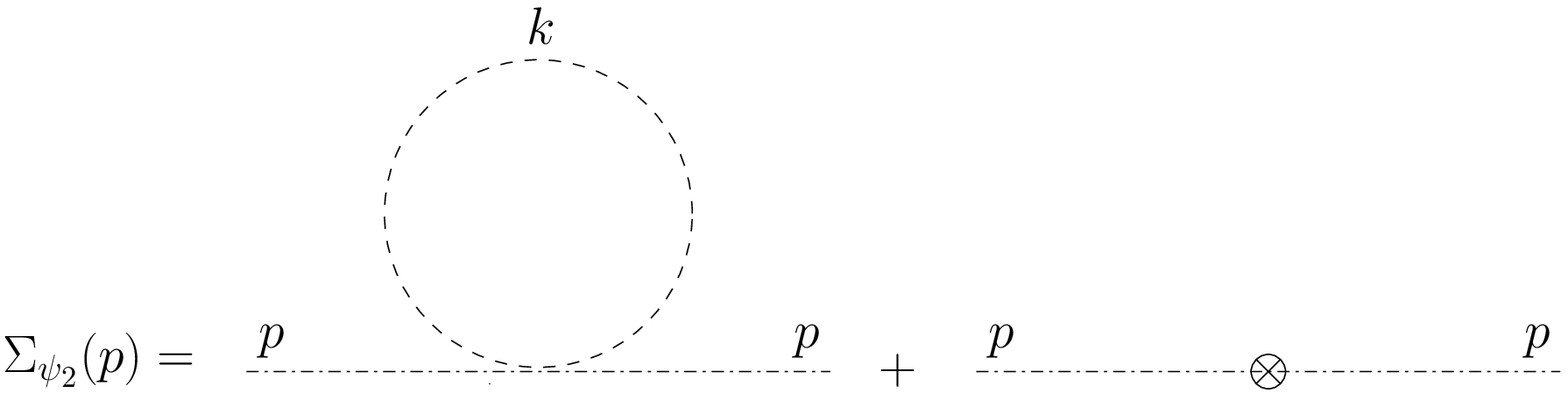}
\caption{\label{fig:self-energy-psi2}
 Self-energy contribution for the physical fermion $\psi_{2}$.
 The dashed propagator corresponds to the mediator scalar field $\sigma$
 and the counterterm is given in Fig.~\ref{fig:counterterm-psi2}.}
\vspace*{15mm}
\includegraphics[width=0.45\linewidth]{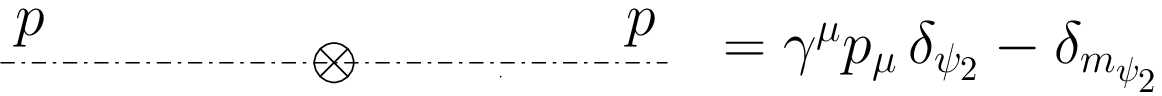}
\caption{\label{fig:counterterm-psi2}
Counterterm for Fig.~\ref{fig:self-energy-psi2}.}
\end{figure}

The self-energy for the $\psi_{2}$ fermion including the counterterm is given by Figs.~\ref{fig:self-energy-psi2} and \ref{fig:counterterm-psi2}.
\vspace{0cm}
\noindent
It is now clear that the $\psi_{2}$ self-energy
of Fig.~\ref{fig:self-energy-psi2} has the same structure
as the $\phi$ self-energy of Fig.~\ref{fig:self-energy-phi}.
In other words, the divergent integral for the $\psi_{2}$ self-energy is
proportional to \eqref{eq:divergent-integral-scalar}.

Consider again Pauli--Villars regularization
and take over the relevant results from Sec.~\ref{subsec:Scalar-field-phi}.
The counterterm is then given by
\begin{subequations}
\begin{eqnarray}
\delta_{\psi_2} &=& 0   \,,
\\[2mm]
\delta_{m_{\psi_2}}
&=&
g_{f,2}\,\lambda\,\frac{\Lambda^2}{\left(4\pi\right)^2} \left(1-\log\frac{\Lambda^2}{m_\sigma^2}\right)   \,,
\end{eqnarray}
\end{subequations}
and, adapting the constants in \eqref{eq:massphi-PV-reg},
the following generated mass is obtained:
\beq
M_{\psi_{2}}\,\Big|^\text{(PV\,reg.)}
=\lim_{p\rightarrow 0} \, \Sigma_{\psi_{2}}(p)\,\Big|^\text{(PV\,reg.)}
=g_{f,2}\;\lambda\;\frac{m_{\sigma}^2}{(4\pi)^2}\,,
\eeq
with $m_{\sigma}^2=4\,\lambda^2\,\,\rho_{d}$ from \eqref{massb}.
The point-splitting and dimensional-regularization methods give a
similar result for the generated fermion mass, with the same parametric
dependence $g_{f,2}\,\lambda\,m_{\sigma}^2$.

As the $\psi_{2}$ interaction term of the effective
action \eqref{gasdelta11} involves a factor $\sigma^2$
[instead of the single factor $\sigma$ of the $\psi_1$ interaction term],
the axial transformation \eqref{eq:axial-transformations},
with $\psi_1$ replaced by $\psi_{2}$,
no longer leaves the action \eqref{gasdelta11} invariant.
Hence, there is no axial symmetry for the  $\psi_{2}$ field
to exclude the appearance of a $\psi_{2}$ mass term.

\section{Discussion}
\label{sec:Discussion}

It has become clear over the last years that, if a ``quantum
spacetime foam'' somehow results in a effective classical
spacetime manifold with small-scale structure, this effective
manifold must be Lorentz-invariant to high
precision (at the $10^{-15}$ level in the photon sector~\cite{KS2008}).
In the present article, we have, therefore, investigated a
model of spacetime defects which is Lorentz-invariant over large enough
spacetime volumes.
Even though there is no apparent Lorentz violation in this model,
there may still be nontrivial effects for the propagation of particles.
The quantities that feel the effects of this small-scale structure
must be themselves Lorentz-invariant, an obvious example being mass.
Indeed, we have found a generated mass for both a
scalar field and a Dirac fermion field,
as long as there is no effective axial symmetry
of the toy model considered.
It is, in fact, possible to keep the Dirac fermion field massless,
if an effective axial symmetry is built into the toy model.

Incidentally, the mass generation found here is not entirely surprising,
as mass generation is known to occur for non-Minkowskian
manifolds~\cite{FordYoshimura1979}.
The manifold considered in Ref.~\cite{FordYoshimura1979} has
nontrivial topology
at large length scales (manifold $\mathbb{R}^3\times S^1$),
whereas we are interested in nontrivial topology
at small length scales (cf. the discussion in
Ref.~\cite{KlinkhamerRupp2004}).

Assuming our results to apply to the Higgs scalar boson
of the standard model of elementary particle physics
(with a Higgs mass around $125\,\text{GeV}$~\cite{PDG2016}),
we have from \eqref{eq:massphi-PV-reg} the following upper bound:
\beq\label{eq:lambda2-rhod-bound-from-Higgs}
g_{s}\,\frac{4\,\lambda^2\,\,\rho_{d}}{(4\pi)^2}
\ll
(125\,\text{GeV})^2\,,
\eeq
where the defect density $\rho_{d}$ is defined by \eqref{Pois}
and the dimensional coupling constant $\lambda$ and
the positive dimensionless coupling constant $g_s$
are defined by \eqref{eq:action-gas-delta-functions}.
We have used a strong inequality in
\eqref{eq:lambda2-rhod-bound-from-Higgs},
in order to make sure that the spacetime defects, if present,
do not upset the standard Higgs mechanism for mass generation of
gauge bosons and fermions.

In Sec.~\ref{subsec:Scalar-field-phi}, we already mentioned the possibility that a single energy scale $E_\text{foam}$ controls the
small-scale structure of spacetime and, therefore,
sets the parameters of our model \eqref{eq:action-gas-delta-functions},%
\bsubeqs\label{eq:lambda-rhod-Efoam}
\beqa
\lambda  &\sim& 1/E_\text{foam}\,,
\\[2mm]
\rho_{d} &\sim& (E_\text{foam})^4\,.
\eeqa
\esubeqs
With the scenario \eqref{eq:lambda-rhod-Efoam} and $g_{s}\sim 10^{-2}$,
bound \eqref{eq:lambda2-rhod-bound-from-Higgs} gives
\beq\label{eq:Efoam-bound-from-Higgs}
E_\text{foam}
\ll
8\;\text{TeV}\;\left(\frac{10^{-2}}{g_{s}}\right)^{1/2}\,.
\eeq
In this case, the picture is that the spacetime defects
have an effective size
[of order $1/m_{\sigma}\sim \lambda^{-1}\,(\rho_{d})^{-1/2}$]
and typical distance between neighboring defects
[of order $(\rho_{d})^{-1/4}$]
with the same order of magnitude, $1/E_\text{foam}$.
This single length scale is, however, very much
larger than the Planck length
$1/E_P \approx 1.62 \times 10^{-35}\,\text{m}$,
with $E_P \equiv G^{-1/2} \approx 1.22 \times 10^{16}\,\text{TeV}$.

Scenarios different from \eqref{eq:lambda-rhod-Efoam} are also possible.
One scenario has, for example, $\lambda \sim 1/E_P$
and $\rho_d \ll (10^{-2}/g_{s})\,(8\;\text{TeV})^2\,(E_{P})^2$.
Ultimately, only the derivation of our model
\eqref{eq:action-gas-delta-functions-epsilon-n-Ppls-Pminus}
from the underlying spacetime (assuming the toy model to be
relevant at all) can decide between the different scenarios.

From a general perspective, it may be of interest to have
found another possible origin of mass, barring questions of
naturalness and the unknown nature of quantum spacetime.
The toy model considered here is rather simple in
that it only gives mass to scalars and Dirac fermions.
More difficult would be spacetime-defect
mass generation for the Weyl fermions and the gauge bosons
of a chiral gauge theory (the type of
theory relevant to the standard model).
For the Weyl fermions, we may consider replacing
the single real scalar field $\sigma(x)$ of our model
by a complex scalar field $\Sigma(x)$  in
an appropriate representation of the gauge group and
using this scalar $\Sigma$
in a gauge-invariant Yukawa-type coupling term.
For the gauge bosons,
perhaps the spacetime-defect mechanism can be
merged with a modified version of the Higgs mechanism.

\section*{\hspace*{-4.5mm}ACKNOWLEDGMENTS}
\vspace*{-0mm}\noindent
We thank the referee for useful comments.

\begin{appendix}

\section{Random-phase approximation}
\label{app:Random-phase-approximation}

Let us investigate a double momentum integral of the following form:
\beq
\int d^4 p\, d^4 q \;f(p,\,q)\;J(p)\,J^\star(q)
\label{genint}\,,
\eeq
where $J(p)$ is defined by
\beq
J(p)\equiv \sum_{n} \epsilon_{n} \, e^{i p x_{n}}\,,
\eeq	
with random defect positions $x_{n} \in \mathbb{R}^4$
and random defect charges $\epsilon_{n} \in \{ -1,\, +1\}$
as discussed in Secs.~\ref{sec:Poisson} and \ref{subsec:General-remarks},
respectively. The defect index $n$ runs over positive integers,
\beq\label{eq:mathcalN}
n \in \mathcal{N}\equiv \{ 1,\,2,\,3,\, \ldots ,\,N\}\,,
\eeq
where $N$ is taken to infinity at the end of the calculation,
together with the volume $V_{4,\,\text{cutoff}}$ discussed in Sec.~\ref{subsec:Massless-mediator}.

Now consider the product of two $J$'s at different momenta,
\beq\label{eq:product-Jp-Jqstar}
J(p)\,J^\star(q)= \sum_{m,n}\epsilon_{m}\,\epsilon_{n}\,
e^{i p x_{n}-iqx_{m}}\,,
\eeq
where the indices $m$ and $n$ run over the set
$\mathcal{N}$ as given by \eqref{eq:mathcalN}.
The product \eqref{eq:product-Jp-Jqstar} can be split in two parts,
a single sum and a double sum,
\beq
J(p)\,J^\star(q)=\sum_{n}e^{i x_{n} (p-q)}+\sum_{m\neq n}\epsilon_{m}\,\epsilon_{n} \, e^{i p x_{n}-iq x_{m}}
\label{current1}\,.
\eeq
It is instructive to
rewrite the double sum of (\ref{current1}) as follows
\beq
\sum_{m\neq n}\epsilon_{n}\,\epsilon_{m} \, e^{i p x_n-iqx_m}
=\sum_{m\neq n} e^{i p x_n-iqx_m+i \pi\,g_{m,n}}\,,
\label{eq3}
\eeq
with
\beq
  g_{m,n} \equiv
\frac{1}{2}\;(1-\epsilon_{n}\,\epsilon_{m}) \in \{ 0,\, 1\} \,.
\eeq
Hence, the double sum \eqref{eq3} is a sum of pure random phases,
even for the case $p\rightarrow q$.
For $p=q=0$, the norm of this double sum is of order $\sqrt{N}$,
and the same holds for generic values of $p$ and $q$.
The norm of the single sum of (\ref{current1}) has a maximum value
$N$ for $p=q$.

Let us briefly recapitulate.
As discussed in Sec.~\ref{subsec:General-remarks}, the charges of individual defects are chosen randomly and the average charge approaches
zero in the limit of an infinite number of defects
($N\to\infty$).
The random distribution of defect charges $\epsilon_n$
also makes that the double sum on the right-hand side of
(\ref{current1}) scatters around zero for a large number of defects
($N\gg 1$), with a spread proportional to $\sqrt{N}$.
Note that the single sum on the right-hand side of (\ref{current1})
is independent of the charge  distribution, being
proportional to $\epsilon_n^2=1$, and takes the value $N$ for $p=q$.
For further discussion and a numerical example, see Sec.~IV of
Ref.~\cite{KlinkhamerRupp2004}.

With the double sum on the right-hand side of
(\ref{current1}) being subdominant,
we can approximate the product (\ref{eq:product-Jp-Jqstar}) by
\beq
J(p)\,J^\star(q)\approx \sum_{n}e^{i x_{n} (p-q)}
\label{current2}
\,,
\eeq
where the sum approaches the value $N$ as $p \to q$.
If we now take into account that the points $x_{n}$ are randomly distributed by a Poisson sprinkling process, there is only a significant contribution to  (\ref{current2}) if $p\approx q$.
This allows us to replace the sum (\ref{current2}) by a delta function, which becomes exact under the identification (\ref{assump}).
We, finally, have
\beq
J(p)\,J^\star(q)\approx \rho_{d} \int d^4 z \,e^{i z (p-q)}
=(2\pi)^4 \,\rho_{d}\;\delta^{(4)}(p-q)\,,
\eeq
which simplifies the original double integral \eqref{genint}.

\end{appendix}


\end{document}